\documentstyle[aps,12pt,epsfig,manuscript,amsbsy]{revtex}
\global\firstfigfalse
\begin{document}
\preprint{INJE-TP-99-6}
\def\overlay#1#2{\setbox0=\hbox{#1}\setbox1=\hbox to \wd0{\hss #2\hss}#1%
\hskip -2\wd0\copy1}

\title{Noncummutative Geometry and Anyonic Field Theory in the 
Magnetic Field}

\author{ Y.S. Myung and H.W. Lee}
\address{Department of Physics, Inje University, Kimhae 621-749, Korea}

\maketitle

\begin{abstract}
We consider an easy way to get the noncommutative spacetime in 
Minkowski space. 
This corresponds to introducing a magnetic field 
${\rm\bf B} = B \hat {\rm\bf k}$ in the plane.
We construct a green's function in coordinate space which includes 
a Moyal phase factor.
The projection to the lowest Landau level(LLL) is necessary for a 
simple calculation.
Using this green's function and a second quantized formalism, 
we study the thermodynamic property of 
the anyons on the noncommutative geometry.
It turns out that the Moyal phase factors contribute to the
thermodynamic potential $\Omega$ as opposed to the
free-particle nature.
\end{abstract}

\newpage
\section{Introduction}
\label{introduction}
Gauge theories on noncommutative space are relevant to the 
quantization of D-branes in background $B_{\mu\nu}$ 
fields\cite{Ala98JHEP02003,Mic98JHEP02008,She99PLB119,Mal9908134,Sei9908142}.
The effect of noncommutativity is given in the momentum space 
vertices as the from of Moyal bracket phase.
To derive this factor, the authors in Ref.\cite{Big9908056} consider a 
dipole in a magnetic field {\bf B}.
In the limit of strong magnetic field ({\bf B}$\to \infty$), 
this dipole is frozen into the 
lowest Landau level(LLL).
Interaction of such dipoles include the Moyal bracket phase factor
of $e^{i p \wedge q}$ with $p \wedge q = \epsilon^{ij} p_i q_j / B$.
However, in this process, the dipole in a strong magnetic field is 
turned out to be a galileian particle of mass $M$ because they neglect 
the kinetic terms.

It is important to note that the noncommutativity comes just from the 
presence of the magnetic field.
The condition of ``strong'' makes the calculation 
easy and does not matter with the noncommutativity.
In this sense it is valuable to study a field theory both in the presence 
of a magnetic field and in the coordinate space.
Actually one can derive the factor of $e^{-i \tau \wedge \tau'}$ from the 
operation ${T}_{\boldsymbol{\tau}}{T}_{\boldsymbol{\tau}'}
{T}_{-\boldsymbol{\tau}}
{T}_{-\boldsymbol{\tau}'}$ of magnetic translation 
operator(${T}_{\boldsymbol{\tau}}$)\cite{Gir9907002}. 
This means that when an electron travels around a parallelogram 
generated by ${T}_{\boldsymbol{\tau}}{T}_{\boldsymbol{\tau}'}
{T}_{-\boldsymbol{\tau}}
{T}_{-\boldsymbol{\tau}'}$ it picks up a phase
$\phi = 2 \pi \Phi/\Phi_0 = B \hat {\rm\bf k} \cdot 
(\boldsymbol{\tau} \times \boldsymbol{\tau}' ) 
\equiv \tau \wedge \tau'$, where $\Phi$ is the magnetic 
flux in the parallelogram and $\Phi_0 = hc/e$ is the flux quantum.
The only difference between the momentum and coordinate spaces is 
the phase factor: in the momentum space it is proportional to $1/B$ 
while in the coordinate space it is proportional to $B$.
This is so because the correct dimensions should be recovered.
If we introduce the green's function in a magnetic field and 
in the coordinate space, this 
noncommutative situation is shown up clearly.
For example, the one particle green's function for a free particle 
is\cite{Vei92NPB715} 
\begin{equation}
G_\beta^{free}({\rm\bf r}_2, {\rm\bf r}_1) = { m \over {2 \pi \beta}}
e^{-m r_{12}^2/2 \beta} ,
\label{green-function}
\end{equation}
whereas the one in the LLL is given by\cite{Kim93PRD4839}
\begin{equation}
G_\beta^B({\rm\bf r}_2, {\rm\bf r}_1) = { {m \omega_c} \over \pi }
e^{-\beta \omega_c} 
\exp \left [ - {{m \omega_c} \over 2} \left \{ r_{21}^2 
+ 2 i \epsilon \hat {\rm\bf k} \cdot ({\rm\bf r}_2 \times {\rm\bf r}_1 ) 
\right \} \right ],
\label{green-LLL}
\end{equation}
with $\omega_c = e |{\rm\bf B} |/2 mc$ and 
$\epsilon = B/|{\rm\bf B} |$.
We note that $G_\beta^{free}$ is symmetric under 
$({\rm\bf r}_2, {\rm\bf r}_1) \to ({\rm\bf r}_1, {\rm\bf r}_2)$.
But $G_\beta^B$ no longer carries such a symmetry 
because of the presence of the 
phase factor.
The phase factor originates from a subtle, combined property of 
translation and gauge transformation 
in the presence of magnetic field({\it i.e.}, the magnetic translation).

In this paper, we study the nonrelativistic anyonic model as a model of 
(2+1)D field theory on the noncommutative geometry.
This model has already introduced to study the fractional 
statistics\cite{Wil90} and 
anyonic physics\cite{Vei92NPB715,Kim93PRD4839}.
Now we reconsider this model to explore its hidden noncommutative property.
We can regard this model as a simple model which shows the noncommutativity. 
Actually the noncommutativity appears as the matrix $M$ in the form of 
the anti-symmetric submatrix($d_{ij}$).

\section{Anyonic Model in a Magnetic Field}
\label{model}
We start with the Lagrangian for an ideal gas of fractional 
particles (nonrelativistic anyons) in a magnetic field
(${\rm\bf B} = B \hat {\rm\bf k}$)\cite{Wil90}
\begin{equation}
{\cal L} = \sum_{i=1}^{N}
\left [ { m \over 2} \dot {\rm \bf x}_i^2 + 
 q \dot {\rm\bf x}_i \cdot {\rm\bf A}_i
+ q \left \{ 
-{\rm\bf a}_0({\rm\bf x}_i) + \dot {\rm\bf x}_i \cdot {\rm\bf a}_i \right \} 
 \right ]
+ {1 \over {2 \alpha}} \int d^2 x \epsilon_{\rho \sigma \tau}
{a}_\rho \partial_\sigma {a}_\tau ,
\label{lagrangian}
\end{equation}
where ${\rm\bf x}_i$ is the $i$th particle coordinates, $q$(charge= $-e$), 
${\rm\bf A}_i= ( - {B \over 2} y_i, { B \over 2} x_i, 0 ) $
with $\nabla_i \times {\rm\bf A}_i = {\rm\bf B}$,
${\rm\bf a}_i$(statistical gauge potential)  and 
${{\rm\bf a}}_0$(scalar potential). 
The first term is the kinetic term for nonrelativistic particles.
The second one is their interaction with a magnetic field.
The third term is their coupling with the statistical gauge potential.
The last term is just the Chen-Simons term which associates 
with each particle fictitious flux $\alpha q$.
$\alpha$ plays a role of the statistical parameter.
The anyons are considered as identical particles (fermions or 
hard-core bosons) with the flux $\alpha q$.
On later, we need to introduce a harmonic potential term of 
($- \sum_{i=1}^N {1 \over 2} m \omega^2 r_i^2$) to regularize 
the divergences\cite{Vei92NPB715,Kim93PRD4839}.
Then our model (\ref{lagrangian}) is very similar to (3) 
of ref.\cite{Big9908056}.
The difference is that in our case all particles carry the same 
charge $q = -e$, but Bigatti and Susskind considered a 
dipole with harmonic interaction between the charges to connect 
the string theory.
Further their way to lead the Moyal phase factor is artificial.
However we include this factor into the green's function without 
any vertex correction.
Using this green's function, we study the anyons on the 
noncommutative geometry.

After some calculation, one finds the corresponding Hamiltonian as
\begin{equation}
{\cal H} = { 1\over {2 m} } \sum_{i=1}^N 
\left ( \boldsymbol{\pi}_i + {e \over c} {\rm\bf a}_i \right )^2,
\label{hamiltonian}
\end{equation}
where the mechanical momentum $\boldsymbol{\pi}_i$ is given by 
\begin{equation}
\boldsymbol{\pi}_i = {\rm\bf p}_i + { e \over c} {\rm\bf A}_i
\label{momentum}
\end{equation}
with the canonical momentum ${\rm\bf p}_i$.
And the statistical gauge potential ${\rm\bf a}_i$ takes the form
\begin{equation}
{\rm\bf a}_i = - \alpha { e \over c} \sum_{i \ne j}^N 
{{ \hat {\rm\bf k} \times {\rm\bf r}_{ij}} \over {r_{ij}^2} }
\label{gaugepotential}
\end{equation}
which satisfies the Coulomb gauge condition 
${\nabla}_i \cdot {\rm\bf a}_i =0$ and 
$\nabla_i \times {\rm\bf a}_i \equiv  b\hat {\rm\bf k}$.
Here we set $\hbar = 1$ and $\hat {\rm\bf k}$ is the unit vector perpendicular 
to the plane.
Let us see how the noncommutativity comes out from the presence of a 
magnetic field.
In the absence of a magnetic field, the commutator of the 
momentum $\boldsymbol{\pi}_i$ is given by
\begin{equation}
\left [ {\pi}_i^x, {\pi}_i^y \right ]_{B=0} = 0.
\label{commutator}
\end{equation}
But in the presence of a magnetic field  the commutator leads to
\begin{equation}
\left [ {\pi}_i^x, {\pi}_i^y \right ]_{B \ne 0} = 
i { e \over c} B \delta_{ij}.
\label{commutatorB}
\end{equation}
This is an easy way to get a noncommutative spacetime in the plane.

The Schr\"odinger equation for the $N$ anyon is
\begin{equation}
H \Psi({\rm\bf r}_1, ... , {\rm\bf r}_N ) = 
   E \Psi({\rm\bf r}_1, \cdots, {\rm\bf r}_N ).
\label{schrodinger}
\end{equation}
We now treat the
$\alpha$- and $\alpha^2$-anyonic interactions in (\ref{hamiltonian}) as the 
perturbations of the Hamiltonian $H^0$:
\begin{eqnarray}
                                            H&=&H^{0}+ \Delta H ,  
\label{H} \\
        H^{0}&=& \sum_{i=1}^N { {\boldsymbol{\pi}}_i^2 \over 2m} ,
\label{H0} \\
   \Delta H &=& \sum_{i=1}^N {e \over 2mc} 
\left \{ \left ( {\rm\bf p}_i + {e \over c} {\rm\bf A}_{i} \right ) \cdot 
   {\rm\bf a}_{i}+ {\rm\bf a}_{i} \cdot \left ( {\rm\bf p}_{i} + 
    {e \over c} {\rm\bf A}_{i} \right )  
          +{e \over c} {\rm\bf a}_{i} \cdot {\rm\bf a}_{i} \right \} .  
\label{DeltaH}
\end{eqnarray}
Here $H^{0}$ describes $N$ particles (bosons or fermions) moving in 
the uniform magnetic field. 
As it stands, the model with ${\cal H}^0$ is very important.
In particular, the study of (2+1)D nonrelativistic fermions in the presence of 
magnetic field is relevant to the 
fractional quantum Hall effect(FQHE)\cite{Gir9907002}.
Such a system has a further connection with (1+1)D $c=1$ 
string model\cite{Iso92PLB143,Cap93PLB100}.
That is, a system of (2+1)D nonrelativistic fermions in the LLL is dual to 
a boundary system of (1+1)D nonrelativistic fermions 
($c=1$ string model).
This is very similar to the AdS$_3$/CFT correspondence in the sense 
of the bulk/boundary dynamics\cite{Myu99PRD044028}.

There exists a conceptual difficulty in doing the perturbation 
near $\alpha =0$.
Because of the singular nature of the $\alpha^{2}$-interaction 
\begin{equation}
 { \alpha^2 \over 2m} \sum_{i=1}^N 
\left \{ 2 \sum_{i<j}^N {1 \over r_{ij}^2}+ \sum_{i \ne k,l(k\ne l)}^N
      {{( {\rm\bf k} \times {\rm\bf r}_{ik}) \cdot ( {\rm\bf k} \times {\rm\bf r}_{il}) }
      \over {r_{ik}^2 r_{il}^2} } \right \}
\label{singular}
\end{equation}
and the fact that a wave function does not 
vanish when any two bosonic particles approach
each other ($r_{ij} \to 0$), a naive perturbation 
would lead to an infinite energy
shift\cite{Gir9907002,Vei92NPB715}. In order to overcome this difficulty, we use
the improved technique of the perturbation. 
The singular nature of the interaction forces the
real wave function to vanish when $r_{ij}\to 0$. 
Hence we redefine the $N$-body wave function as 
$    \Psi ( {\rm\bf r}_1,..., {\rm\bf r}_N) = 
\prod_{i<j} r_{ij}^\gamma \tilde \Psi ( {\rm\bf r}_1,..., {\rm\bf r}_N) 
$.
One can easily show that all divergent terms in (\ref{singular}) 
disappear if $\gamma$ is equal to $| \alpha |$.
It is worth noting that the prefactor 
$\prod_{i<j} r_{ij}^{|\alpha|}$ can be interpreted
as a factor which optimizes the dynamical short-range avoidance between anyons.
The resulting equation is 
$   (H^{0}+ \Delta \tilde H ) \tilde \Psi = E \tilde \Psi 
$.
Here the perturbed Hamiltonian $\Delta \tilde H$ is given by 
\begin{equation}
\Delta \tilde H = \sum_{i<j}^N 
\left \{ i {\alpha \over m} { {\hat{\rm\bf k} \times {\rm\bf r}_{ij}} \over
    {r_{ij}^2}} \cdot ( \boldsymbol{\partial}_i- \boldsymbol{\partial}_j)  
  - { {| \alpha |} \over m} { {\rm\bf r}_{ij} \over { r_{ij}^2}}
  \cdot ( \boldsymbol{\partial}_i- 
      \boldsymbol{\partial}_j) + \alpha \epsilon \omega_c \right \} .
\label{delta-H}
\end{equation}
The interaction terms in (\ref{delta-H}) are two-body
interactions, contrary to (\ref{DeltaH}) 
where three-body interactions are present. This 
approach is based on the quantum-mechanical framework 
at first order in $\alpha$. For second
and higher corrections, this is no longer useful.
Rather, it is appropriate to use a quantum field theory. 

\section{Second Quantized Formalism}
\label{formalism}
In order to compute perturbatively the thermodynamic potential $\Omega$ in the grand canonical
ensemble, we introduce a second quantized formalism (finite-temperature
quantum field theory)\cite{Gir9907002,Vei92NPB715}. 
This formalism is also very useful for representing the magnetic 
translation symmetry.
The thermodynamic potential is given by 
\begin{equation}
 \Omega =- \beta PV =- \ln {\rm Tr} e^{- \beta ( {\cal H} - \mu {\cal N} )} ,
\label{OmegaPV}
\end{equation}
where ${\cal H}$ and ${\cal N}$ stand respectively for 
the second quantized Hamiltonian and the
number operator of anyons, and $\mu$ is the chemical potential 
[ $z = \exp ( \beta \mu )$]. In terms of
a second quantized field $\psi$ the second quantized 
Hamiltonian ${\cal H}$ takes the form 
\begin{equation}
 {\cal H} = {\cal H}^{0}+ {1 \over 2} 
\int d {\rm\bf r}_1 d {\rm\bf r}_2 \psi^\dag ( {\rm\bf r}_1) 
\psi^\dag ( {\rm\bf r}_2) {\cal V} ( {\rm\bf r}_1- {\rm\bf r}_2)  
 \psi ( {\rm\bf r}_2) \psi ( {\rm\bf r}_1) 
\label{calH}
\end{equation}
with 
\begin{equation}
{\cal H}^{0} = { 1\over 2m} \psi^\dag
( {\rm\bf p} + {e \over c} {\rm\bf A} )^2 \psi.
\label{calH0}
\end{equation}
Here ${\cal V}$ is the anyonic interaction given by 
\begin{eqnarray}
 {\cal V} ( {\rm\bf r}_{1}- {\rm\bf r}_{2}) &=& 
  {\cal V} ( {\rm\bf r}_{1}, {\rm\bf r}_{2}) + 
   {\cal V} ( {\rm\bf r}_{2}, {\rm\bf r}_{1}) , 
\nonumber \\ 
 {\cal V} ( {\rm\bf r}_{1}, {\rm\bf r}_{2}) &=& 
    {{i \alpha \hat{\rm\bf k} \times {\rm\bf r}_{12}} 
      \over {m r_{12}^2}} \cdot \boldsymbol{\partial}_1
     - {{| \alpha | {\rm\bf r}_{12}} 
         \over {m r_{12}^2}} \cdot \boldsymbol{\partial}_1 
        + { {\alpha \epsilon \omega_c} \over 2 }.
\label{calV}
\end{eqnarray}
The first term (anyonic vertex) in (\ref{calV}) 
measures the energy change in the nonzero angular
momentum sector and materializes, only in the presence of a 
magnetic field, in the virial
coefficients. The second (short-range improved vertex) 
comes from optimizing the dynamical
short-range avoidance between anyons and measures the energy change in the zero angular
momentum sector. The last (constant vertex) couples the statistical 
parameter $\alpha$ to the
magnetic field and plays a crucial role in cancellation of divergences. It is important to note
that the $| \alpha |$ term is not Hermitian, and is complex as much as the anyonic vertex. 
Instead, we construct the simple Hermitian vertex
\begin{equation}
 {\cal V}^{H}( {\rm\bf r}_{1}, {\rm\bf r}_{2}) = 
   {1 \over 2} \left \{ {\cal V} ( {\rm\bf r}_{1}, {\rm\bf r}_{2}) 
        + {\cal V}^{\dag}( {\rm\bf r}_{1}, {\rm\bf r}_{2})\right \} .  
\nonumber
\end{equation}
The explicit form of this vertex is given by 
\begin{equation}
   {\cal V}^{H}( {\rm\bf r}_{1}, {\rm\bf r}_{2}) = 
    { {i \alpha \hat{\rm\bf k} \times {\rm\bf r}_{12}} 
       \over {m r_{12}^{2}}} \cdot \boldsymbol{\partial}_1
  + {\pi \over m} | \alpha | \delta ( {\rm\bf r}_{12}) 
  + { {\alpha \epsilon \omega_{c}} \over 2}.
\label{calVH}
\end{equation}
The equivalence of the $| \alpha |$-term in 
${\cal V} ( {\rm\bf r}_{1}, {\rm\bf r}_{2})$ and the 
$| \alpha |$-term
in ${\cal V}^{H}( {\rm\bf r}_{1}, {\rm\bf r}_{2})$ 
was confirmed to be valid at second order in $\alpha$\cite{Vei92NPB715}.

We treat $\alpha$ and $| \alpha |$ as small parameters and expand perturbatively the thermodynamic
potential $\Omega$ : 
\begin{equation}
 \Omega = \Omega_0 - \sum_{i=1}^\infty (-1)^i \int_0^\beta d \beta_1
        \int_0^{\beta_1} d \beta_2  \cdots 
 \int_0^{\beta_{i-1}} d \beta_i \langle {\cal V} ( \beta_{1}) 
  {\cal V} ( \beta_{2}) \cdots {\cal V} ( \beta_i) \rangle^{c} ,
\label{Omega}
\end{equation}
where 
\begin{eqnarray}
\Omega_0 &=& \pm \ln {\rm Tr} e^{-\beta ({\cal H}_0 - \mu {\cal N})},
\label{Omega0} \\
   {\cal V}( \beta_1) &=& {1 \over 2} 
    \int d {\rm\bf r}_1 d {\rm\bf r}_{2} 
    \psi^{\dag}( {\rm\bf r}_1, \beta_1)
    \psi^{\dag}( {\rm\bf r}_2, \beta_1) 
    {\cal V} ( {\rm\bf r}_1- {\rm\bf r}_2) 
     \psi ( {\rm\bf r}_2,\beta_1) 
     \psi ( {\rm\bf r}_1, \beta_1) .  
\label{calVbeta1}
\end{eqnarray}
${\cal V}( \beta_{1})$ is the two-anyonic interaction built from the
thermal second quantized field 
\begin{equation}
  \psi( {\rm\bf r}_{1}, \beta_{1}) = 
      e^{\beta_{1}({\cal H}_0 - \mu {\cal N})} 
      \psi ( {\rm\bf r}_{1}) e^{-\beta_{1}( {\cal H}_0 - \mu {\cal N})} .  
\label{2ndfield}
\end{equation}
The upper index $c$ in (\ref{Omega}) means that one omits any disconnected diagram when using the Wick
theorem. The Wick theorem is performed through the one particle thermal green's function [
$\psi^{\dag}( {\rm\bf r}_{1}, \beta_{1}) 
\psi ( {\rm\bf r}_{2}, \beta_{2})$ ]. The thermal propagator in a power
series of $z$ is then 
\begin{equation}
  [ \psi^{\dag}( {\rm\bf r}_{1}, \beta_{1}) 
    \psi ( {\rm\bf r}_2, \beta_2)] = 
     \sum_{s=1,s=0}^\infty (\pm)^{s+1} 
 z^{s-\beta_{12}/ \beta } \langle {\rm\bf r}_2| 
   e^{-(s\beta - \beta_{12}) H} |{\rm\bf r}_1 \rangle ,
\label{thermal}
\end{equation}
where $\beta_{12}= \beta_{1}- \beta_{2}$ and 
$H$ is the one particle Hamiltonian.
Here $\pm$ refers to Bose/Fermi cases. When $\beta_{12} \ge 0$, $s$ 
starts at
$s=1$; whereas when $\beta_{12}<0$, it starts at $s=0$. 
In the lowest Landau level of $B \to \infty$, 
the one-particle green's
function at temperature $s \beta - \beta_{12}$ is : 
\begin{eqnarray}
    G_{s\beta - \beta_{12}}^{B}( {\rm\bf r}_{2}, {\rm\bf r}_{1})&=& 
   \langle {\rm\bf r}_{2}| e^{-(s\beta -
                              \beta_{12}) H_{\rm LLL }}| {\rm\bf r}_{1}\rangle  
\nonumber \\
   &=& {{m \omega_{c}} \over \pi } 
    e^{-(s\beta - \beta_{12}) \omega_{c}} 
    \exp \left [ - {{m \omega_{c}} \over 2}
      \left \{ r_{21}^2+2i\epsilon \hat {\rm\bf k} \cdot 
           ( {\rm\bf r}_{2} \times {\rm\bf r}_{1})
     \right \}  \right ] .
\label{greentemp}
\end{eqnarray}
We here consider the statistical mechanics of a gas of 
anyons in a strong magnetic field, and in
the thermodynamic limit. A naive perturbative calculation of thermodynamic potential consists
in working with (\ref{greentemp}). 
However, the presence of a free-particle nature and a 
phase factor in the exponent of (\ref{greentemp}) leads to
the unwanted result (a divergent quantity). 
A good regularization procedure should be introduced to resolve this problem
by adding an extra potential term of confining nature. 
For simplicity we introduce a harmonic
regulator to give an unambiguous meaning to all 
diagrams(Fig.\ref{first-order} - Fig.\ref{second-cluster}). 
This amounts to adding to (\ref{H0}) a term
of $\sum_{i=1}^{N}{1 \over 2}m \omega^{2} r_i^{2}$ 
and the thermodynamic limit is understood as $\omega \to 0$.
The one-particle green's function at temperature $s \beta$ in 
a constant magnetic field with a harmonic regulator reads 
\begin{eqnarray}
  G_{s \beta }^{\rm full} ( {\rm\bf r}_{2}, {\rm\bf r}_{1}) &=& 
     {{m \omega_t} \over {2 \pi \sinh s \beta \omega_t }}
    \exp \left [
     - { {m \omega_t} \over {2 \sinh s \beta \omega_t}} 
    \left \{ ( \cosh s \beta \omega_c) r_{12}^2
\right . \right .
\nonumber \\
   && ~~~~~
        +( \cosh s \beta \omega_t- \cosh s \beta \omega_c)
            (r_1^2+r_2^2)
    + (2i \epsilon \sinh s \beta \omega_c) \hat{\rm\bf k} 
        \cdot ( {\rm\bf r}_{2} \times {\rm\bf r}_{1})
    \big \} \Big ],
\label{greenfull}
\end{eqnarray}
where $\omega_t= \sqrt { \omega_c^2+\omega^2}$. 
In the lowest Landau level, the regularized
one-particle green's function at temperature $s \beta$ 
is obtained by taking the limit if $\omega_c \to \infty$
and $\omega \to 0$: 
\begin{equation}
  G_{s \beta } ( {\rm\bf r}_{1}, {\rm\bf r}_{2}) = 
   { {m \omega_c} \over \pi}  a_s e^{-s\beta \omega_c} 
    \exp \left [ - { {m \omega_c} \over 2}  
     a_s \left \{ r_{12}^2+2 i \epsilon \hat{\rm\bf k} \cdot 
     ( {\rm\bf r}_{1} \times {\rm\bf r}_{2}) \right \} 
     - b_s (r_1^2+r_2^2) \right ] ,
\label{greensbeta}
\end{equation}
with 
\begin{eqnarray}
\hspace*{-2pt} a_s &=&1 + { \omega^2 \over { 2 \omega_c^2}}(1-s \beta \omega_c) 
     - { \omega^4 \over { 8 \omega_c^4}}
      \left \{ 1+s \beta \omega_c- (s \beta \omega_c)^2 \right \} 
       + { \omega^6 \over {16 \omega_c^6}} \left \{ 1 + s \beta \omega_c-
          {1 \over 3} (s \beta \omega_c)^3 \right \} + \cdots ,  
\nonumber \\
\hspace*{-2pt}
 b_s&=& {{ms \beta \omega^2} \over 4} 
    \left [ 1 + { \omega^2 \over {4 \omega_c^2}}(1-s \beta \omega_c)
   - { \omega^4 \over { 8 \omega_c^4}}
     \left \{ 1- {1 \over 3} (s \beta \omega_c)^2 \right \} \right ] + \cdots .
\nonumber 
\end{eqnarray}
It is sufficient to consider up to the order of $\omega^6$ to obtain the finite results. Care
has to be taken regarding overall normalization. At a given power $s$ of $z$, one has to multiply
the harmonic result by $s$ in order to recover the large volume(area) limit
of $V \to \infty$. 

\section{Moyal Phase Factor and green's function}
\label{moyalphase}
Hereafter we choose $e/c = 1$.
The relevant symmetries of the unperturbed system (\ref{calH0}) 
are the translational and rotational ones.
Here we mainly concern the translational symmetry.
The Hamiltonian ${\cal H}^0$ is invariant under a cocycle transformation 
which is defined through its action over the field operator 
as \cite{Bur91PA281}
\begin{equation}
{\rm U}_{\boldsymbol{\tau}} \psi({\rm\bf r}, t) 
{\rm U}_{\boldsymbol{\tau}}^{-1} =
\exp ( i {\rm\bf A}({\rm\bf r}) \cdot {\boldsymbol{\tau}}) 
\psi({\rm\bf r} - {\boldsymbol{\tau}}, t )  \equiv 
T_{\boldsymbol{\tau}} \psi({\rm\bf r}, t ),
\label{cocycle}
\end{equation}
where ${\rm U}_{\boldsymbol{\tau}}$ is the unitary operator representing 
the translation  in the second quantized Fock space.
Also the above is the defining equation of the magnetic 
translation operator $T_{\boldsymbol{\tau}}$.
From (\ref{cocycle}) the generator(${\rm\bf G}^c$) of 
the cocycle transformation is derived as
\begin{equation}
T_{\boldsymbol{\tau}} \psi({\rm\bf r}, t) =
\exp(-i {\boldsymbol{\tau}} \cdot {\rm\bf G}^c({\rm\bf r})) \psi({\rm\bf r}, t), ~
{\rm\bf G}^c({\rm\bf r}) = {\rm\bf p} - {\rm\bf A}({\rm\bf r}).
\label{translation}
\end{equation}
Also from (\ref{calH0}), ${\rm\bf G}^c$ is compared with the 
momentum $\boldsymbol{\pi} = {\rm\bf p} + {\rm\bf A} $. 
For our purpose let us introduce the complex coordinates as
\begin{equation}
z = \sqrt{{{|{\rm\bf B}|} \over 2}} ( x + i y), ~
\bar z = \sqrt{{{| {\rm\bf B} |} \over 2}} ( x - i y).
\label{coordinate}
\end{equation}
The the Hamiltonian differential operator ( $H = \boldsymbol{\pi}^2/2m$) can be 
rewritten as 
\begin{equation}
H = 2 \omega_c \left \{ - { \partial^2 \over {\partial z \partial \bar z}} 
- {\epsilon \over 2} \left ( z { \partial \over {\partial z}} -
{ \partial \over {\partial \bar z}} \bar z \right ) 
+ { 1\over 4} z \bar z \right \}.
\label{h-diff}
\end{equation}
Further one can define two sets of annihilation and creation operators as
\begin{eqnarray}
{\rm G}_x^c + i {\rm G}_y^c &=& -i \sqrt{2{|{\rm\bf B}|}} 
\left ( {\partial \over {\partial \bar z}}  + { \epsilon \over 2} z \right ) 
\equiv  -i \sqrt{2 {|{\rm\bf B}|}} b,
\label{xpy} \\
({\rm G}_x^c + i {\rm G}_y^c )^\dag &=& {\rm G}_x^c - i {\rm G}_y^c = 
-i \sqrt{2{|{\rm\bf B}|}} 
\left ( {\partial \over {\partial z}}  - { \epsilon \over 2} \bar z \right ) 
\equiv  i \sqrt{2 {|{\rm\bf B}|}} b^\dag,
\label{xmy} \\
\pi_x - i \pi_y &=& -i \sqrt{2{|{\rm\bf B}|}} 
\left ( {\partial \over {\partial z}}  + { \epsilon \over 2} \bar z \right ) 
\equiv  -i \sqrt{2 {|{\rm\bf B}|}} a,
\label{xcpyc} \\
(\pi_x - i \pi_y)^\dag &=& \pi_x + i \pi_y = 
-i \sqrt{2{|{\rm\bf B}|}} 
\left ( {\partial \over {\partial \bar z}}  - { \epsilon \over 2} z \right ) 
\equiv  i \sqrt{2 {|{\rm\bf B}|}} a^\dag,
\label{xcmyc} 
\end{eqnarray}
Here $a(a^\dag)$ is an annihilation(creation) 
operator which mixes the Landau levels. 
On the other hand, $b(b^\dag)$ is an annihilation(creation) 
operator within each Landau level.
The commutation relations are given by
\begin{equation}
[a, a^\dag] =\epsilon, ~ [ b, b^\dag ] = \epsilon,
\label{a-commutator}
\end{equation}
with all other commutators vanishing.
The Hamiltonian operator in (\ref{h-diff}) can be expressed in terms of 
these operators as 
\begin{equation}
H = 2 \omega_c ( a^\dag a + {1 \over 2} ).
\label{h-a}
\end{equation}
The generator of the cocycle transformation in (\ref{translation}) takes 
the following form:
\begin{eqnarray}
&&T_{\boldsymbol{\tau}} \psi(z, \bar z, t) = 
\exp \left \{ \sqrt{{{| {\rm\bf B} | } \over 2 }} ( \tau b^\dag - \tau^* b ) \right \} \psi(z, \bar z, t),
\label{gen-cocycle} \\
&& \tau = \left ( \tau_x + i \tau_y \right ), ~
\tau^* =  \left ( \tau_x - i \tau_y \right ).
\label{TTstar}
\end{eqnarray}
We also have $T_{\boldsymbol{\tau}}T_{\boldsymbol{\tau}'}T_{-\boldsymbol{\tau}}T_{-\boldsymbol{\tau}'} =
e^{-i \tau \wedge \tau'} $ with $\tau \wedge \tau' = { 1 \over l^2 } 
\hat {\rm\bf k} \cdot ( {\tau} \times {\tau}' )$.
Here $l^2$ is a square of magnetic length defined by 
$1/l^2 = \epsilon |{\rm\bf B}| = B$.
This is a familiar feature of the group of translations in a magnetic 
field, because ${\tau} \wedge {\tau}'$ is 
exactly the Moyal phase generated by the 
flux in the parallelogram of 
$\boldsymbol{\tau}$ and $\boldsymbol{\tau}'$ plane.
Hence $T$'s form a ray representation of the magnetic 
translation group.
In fact $T_{\boldsymbol{\tau}}$ translates the particle a distance 
$\hat {\rm\bf k} \times {\boldsymbol{\tau}}$.
This means that different components of $T_{\boldsymbol{\tau}}$ do not 
commute. That is, 
$T_{\boldsymbol{\tau}}T_{{\boldsymbol{\tau}}'} = e^{-i \tau \wedge \tau'} 
T_{{\boldsymbol{\tau}}'}T_{\boldsymbol{\tau}}$.
How does the green's function accomodate this phase factor?
Introducing the flux $\Phi$ enclosed in the parallelogram and $\Phi_0$
(flux quantum), then these take the forms as
\begin{eqnarray}
\Phi &=& {\rm\bf B} \cdot ( {\rm\bf r}_2 \times {\rm\bf r}_1 ),
\label{flux} \\
\Phi_0 &=& {{2 \pi \hbar c} \over { e}} = { 2 \pi}
\label{flux-quanta}
\end{eqnarray}
with $\hbar={e \over c} = 1$.
Then the phase $\phi= 2 \pi \Phi / \Phi_0$ leads to 
$B \hat {\rm\bf k} \cdot ({\rm\bf r}_2 \times 
{\rm\bf r}_1) = { 1 \over l^2} \hat {\rm\bf k} \cdot ({\rm\bf r}_2 \times 
{\rm\bf r}_1) \equiv r_2 \wedge r_1$.
Finally the green's function in a magnetic field is given by 
\begin{equation}
G_\beta^B({\rm\bf r}_2, {\rm\bf r}_1) =
{{ m \omega_c} \over \pi} e^{-\beta \omega_c} 
\exp \left ( - {{ m \omega_c r_{21}^2 } \over 2 } - 
i {{r_2 \wedge r_1} \over 2 } \right ).
\label{green-mag}
\end{equation}
The Moyal phase factor $r_2 \wedge r_1$ in the coordinate space appears 
correctly in the green's function. 
The factor of $1/2$ can be understood from the relation of
$T_{\boldsymbol{\tau}}T_{\boldsymbol{\tau}'} = 
T_{\boldsymbol{\tau}+\boldsymbol{\tau}'} e^{ -i \tau \wedge \tau' /2 } $.
If one uses this green's function to calculate any physical quantity, 
the effect of noncommutativity is taken into account automatically.
This can be done solely by the green's function without 
manipulating vertices as in string theories\cite{Big9908056}.

\section{Structure of Perturbation Theory}
\label{structure}
Here we observe the effect of the Moyal phase factors on the calculation 
of thermodynamic quantities.
For simplicity, we choose the hermitian point vertex
${\cal V}^H({\rm\bf r}_1, {\rm\bf r}_2) = {\pi \over m} |\alpha | 
\delta({\rm\bf r}_{12})$.
This vertex minimizes the divergence problem in higher order corrections.
And this represents the average anyonic effect.
\subsection{First-order calculation}
\label{firstorder}
\begin{center}
\begin{figure}
\epsfig{file=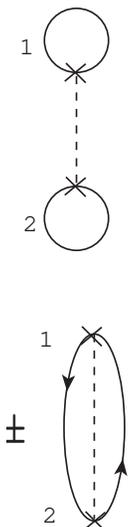, height=7.0cm, clip=}
\caption{
The first-order diagrams. The solid lines (the dashed lines) denote 
the thermal propagator of (\ref{thermal}) (the vertices).
The $\pm$ signs refer to Bose/Fermi cases. 
The arrow($\to$) represents the direction of propagation and 
$\times$ the point of interaction.
}
\label{first-order}
\end{figure}
\end{center}

We are now in a position to derive the first-order correction to $\Omega_0$ 
in (\ref{Omega0}).
One has to consider the two diagrams in Fig.\ref{first-order}.
They correspond to two possible contractions in the Wick expansion
\begin{eqnarray}
\Omega_1 &=& \sum_{s,t \ge 1} (\pm z)^{s+t} \int_0^\beta d \beta_1 
\int d{\rm\bf r}_1 d{\rm\bf r}_2 {\cal V}^H({\rm\bf r}_1, {\rm\bf r}_2)
\nonumber \\
&&\times \left \{ 
G_{s\beta}({\rm\bf r}_1, {\rm\bf r}_1) G_{t\beta}({\rm\bf r}_2, {\rm\bf r}_2)
\pm 
G_{s\beta}({\rm\bf r}_1, {\rm\bf r}_2) G_{t\beta}({\rm\bf r}_2, {\rm\bf r}_1)
\right \}.
\label{omega1}
\end{eqnarray}
\noindent
The first term corresponds to the two-tadpole diagram and the 
second is the conventional diagram.
Considering the point vertex of 
${\pi \over m} |\alpha | \delta({\rm\bf r}_{12})$, 
two terms lead to the same expression.
Hence, in the first-order correction, the phase factors never 
contribute to the thermodynamic potential $\Omega_1$.
$\Omega_1$ takes the form in the large $x$-limit(strong 
magnetic field and low temperature limits)
\begin{equation}
\Omega_1 = | \alpha | {V \over \lambda^2} 
{{1\pm 1} \over 2} 4 x^2 \left [
{{\pm ze^{-x}} \over {1 \mp z e^{-x}}} \right ]^2 ,
\label{large-limit}
\end{equation}
where $\lambda^2 = 2 \pi \beta / m$(thermal 
wavelength), $x = \beta \omega_c$, and $\pm$ refer to 
Bose/Fermi cases.
The equation of state is
\begin{equation}
\beta p V = { V \over \lambda^2} \left [
\pm 2 x \ln(1 \pm \nu_\pm) + 2(1\pm 1) | \alpha | x^2 \nu_\pm^2 
\right ],
\label{state}
\end{equation}
where the filling fraction coefficients($\nu_\pm$) are given by
\begin{equation}
\nu_\pm = { N \over V} \Bigg / \left ( { { e B} \over c } \right ) =
{ {\rho \lambda^2} \over {2 x} } = {{ z e^{-x} } \over { 1 \mp z e^{-x}}}.
\label{filling}
\end{equation}

\subsection{Second-order calculation}
\label{secondorder}
\begin{center}
\begin{figure}
\epsfig{file=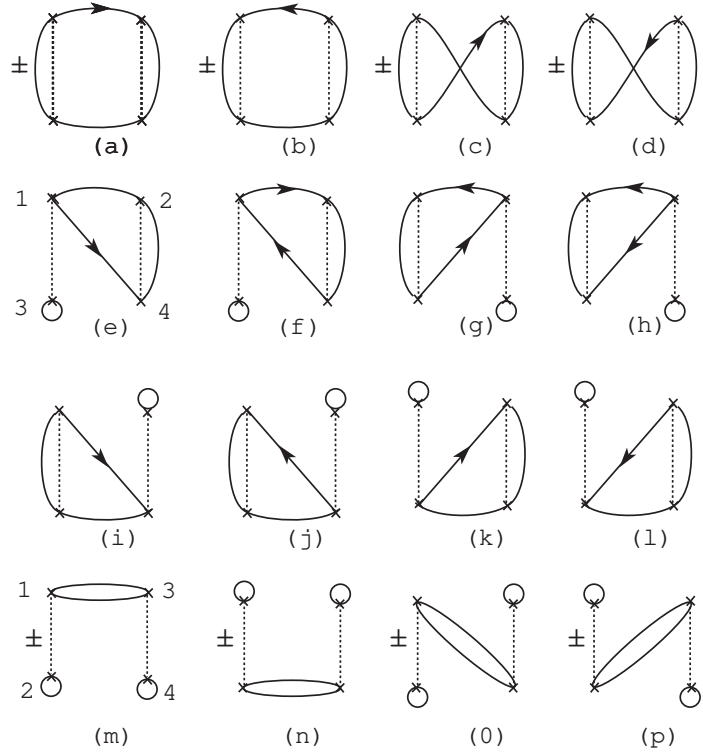, height=10.0cm, clip=}
\caption{
The sixteen second-order diagrams which contribute to the third 
cluster coefficient.
}
\label{third-cluster}
\end{figure}
\end{center}

\begin{center}
\begin{figure}
\epsfig{file=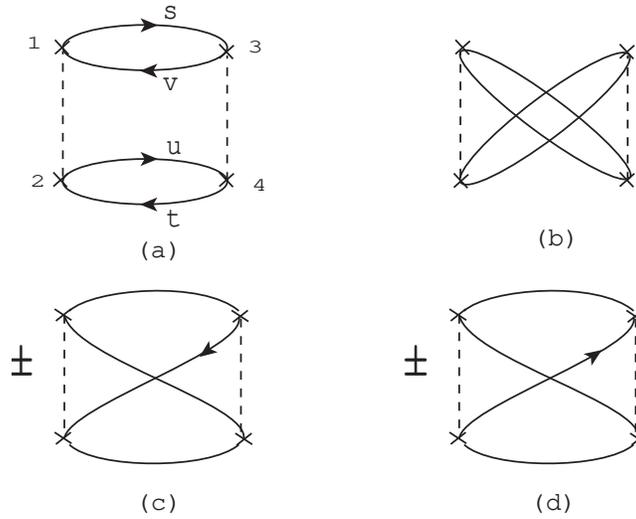, height=7.0cm, clip=}
\caption{
The four second-order diagram which contribute to the second 
and third cluster coefficients.
}
\label{second-cluster}
\end{figure}
\end{center}

We now consider the effect of Moyal phase factors on the second-order 
calculation.
Using the Wick's theorem, one obtains the twenty connected 
diagrams in Fig.\ref{third-cluster} and Fig.\ref{second-cluster}.
\noindent
Each graph with the hermitian vertex  can be computed easily using 
the regularized green's function of (\ref{greensbeta}).
We start with two-tadpoles diagrams.

(1) Diagrams with two tadpoles\\
We consider the diagram of Fig.\ref{third-cluster}(m). 
Applying the Feynman rules, we have
\begin{eqnarray}
\Omega_2^{\rm Fig.\ref{third-cluster}(m)} &=&
\sum_{s,t, u\ge 1; v \ge 0} (\pm z)^{s+t+u+v} \int_0^\beta d \beta_1 
\int_0^{\beta_1} d \beta_2 \int \left ( \prod_{i=1}^4 d{\rm\bf r}_i \right )
{\cal V}^H ({\rm\bf r}_1,{\rm\bf r}_2)
{\cal V}^H ({\rm\bf r}_3,{\rm\bf r}_4)
\nonumber \\
&& \times G_{s \beta}({\rm\bf r}_2,{\rm\bf r}_2)
G_{u \beta}({\rm\bf r}_4,{\rm\bf r}_4)
G_{v\beta+\beta_{12}}({\rm\bf r}_1,{\rm\bf r}_3)
G_{t\beta-\beta_{12}}({\rm\bf r}_3,{\rm\bf r}_1).
\label{omega2m}
\end{eqnarray}
Using the representation of Eq. (\ref{greensbeta}), this reads as 
\begin{eqnarray}
\Omega_2^{\rm Fig.\ref{third-cluster}(m)} &=&
\sum_{s,t, u\ge 1; v \ge 0} (\pm ze^{-x})^{s+t+u+v} \int_0^\beta d \beta_1 
\int_0^{\beta_1} d \beta_2 \int \left ( \prod_{i=1}^4 d{\rm\bf r}_i \right )
a_s a_t a_u a_v \left ( {\omega_c \over \pi} \right )^4 
\nonumber \\
&& 
\times \pi^2 | \alpha |^2 \delta({\rm\bf r}_{12}) \delta({\rm\bf r}_{34})
\exp\left \{
-2 b_s r_2^2 - 2 b_u r_4^2 - {\omega_c \over 2} a_v \left ( r_{13}^2 +
2 i \epsilon \hat {\rm\bf k} \cdot ({\rm\bf r}_1 \times {\rm\bf r}_3 )
\right ) 
\right .
\nonumber \\
&&
\left .
- b_v ( r_1^2 + r_3^2 ) 
- {\omega_c \over 2} a_t \left ( r_{31}^2 +
2 i \epsilon \hat {\rm\bf k} \cdot ({\rm\bf r}_3 \times {\rm\bf r}_1 )
\right ) -b_t ( r_3^2 + r_1^2 )
\right \},
\label{omega2ma}
\end{eqnarray}
where $a_v$ and $b_v$ means $a_{v\beta + \beta_{12}}$ and 
$b_{v\beta + \beta_{12}}$, while $a_t$ and $b_t$ denote
$a_{t\beta - \beta_{12}}$ and $b_{t\beta - \beta_{12}}$.
Here we take $m=1$ for simplicity.
After the integration over the tadpole coordinates ${\rm\bf r}_2$ and 
${\rm\bf r}_4$, the integrand takes the following form:
\begin{equation}
\exp\left\{ - \sum_{i,j=1,3} c_{ij} {\rm\bf r}_i \cdot {\rm\bf r}_j 
- \sum_{i,j=1,3} d_{ij} \hat {\rm\bf k} \cdot 
( {\rm\bf r}_i \times {\rm\bf r}_j )\right \},
\label{integrand}
\end{equation}
where 
\begin{eqnarray}
c_{11} &=& {\omega_c \over 2} (a_v + a_t) + b_v + b_t + 2 b_s ,
\nonumber \\
c_{33} &=& {\omega_c \over 2} (a_v + a_t) + b_v + b_t + 2 b_u,
\nonumber \\
c_{13} &=& -{\omega_c \over 2} (a_v + a_t),
\nonumber \\
d_{13} &=& - i \epsilon {\omega_c \over 2} (a_v - a_t).
\nonumber 
\end{eqnarray}
During the calculation, we define
the matrix $M_4$ 
\begin{equation}
M_4 = \left (
\begin{array}{cccc}
c_{11} & c_{13} & 0 & d_{13} \\
c_{13} & c_{33} & -d_{13} & 0\\
0 & -d_{13} & c_{11} & c_{13} \\
d_{13} & 0 & c_{13} & c_{33} 
\end{array}
\right )
\label{expM}
\end{equation}
The free-particle nature contains in $c_{ij}$, whereas $d_{ij}$ include 
Moyal phase factors.
Performing the gaussian integral over ${\rm\bf r}_1$ and ${\rm\bf r}_3$ 
leads to $\pi^2/ \sqrt{\det M_4}$.
One finds that the determinant of $M_4$ is a perfect square as 
\begin{equation}
\det M_4 = \left ( c_{11} c_{33} - c_{13}^2 - d_{13}^2 \right )^2.
\label{det}
\end{equation}
Here one finds
\begin{equation}
c_{13}^2 + d_{13}^2 = \omega_c^2 a_v a_t,
\label{relation}
\end{equation}
which means that the Moyal phase factor($d_{13}^2$) contributes to the 
thermodynamic potential as opposed to the free-particle 
nature($c_{13}^2$).
This is so because of the pure imaginary of $d_{13}$.
Then, $\Omega_2^{\rm Fig.\ref{third-cluster}(m)}$ takes the form
\begin{eqnarray}
\Omega_2^{\rm Fig.\ref{third-cluster}(m)} &=&
| \alpha | ^2 \omega_c^4
\sum_{s,t, u\ge 1; v \ge 0} (\pm ze^{-x})^{s+t+u+v} \int_0^\beta d \beta_1 
\int_0^{\beta_1} d \beta_2 
a_s a_t a_u a_v 
\nonumber \\
&& \times 
{1 \over { ( A + 2 b_u) ( A + 2 b_s ) - B } }, 
\label{omega2mb}
\end{eqnarray}
where
\begin{equation}
A = { \omega_c \over 2} ( a_v + a_t ) + b_v + b_t , ~ 
B = \omega_c^2 a_v a_t.
\nonumber
\end{equation}
Integrating over the temperatures $(\beta_1, \beta_2)$ followed by the 
summation over $s,t,u$ starting at 1 and over $v$ starting at 0, 
one finds the final contribution in the large $x$-limit
\begin{equation}
\Omega_2^{\rm Fig.\ref{third-cluster}(m)} =
| \alpha |^2 \left [ 
{{ z^3 x^3 V} \over \lambda^2 } 
\right ].
\label{omega2mc}
\end{equation}
The remaining three diagrams of Figs.\ref{third-cluster}(n)-(p) 
contribute to $\Omega_2$ as the same 
form in (\ref{omega2mc}).

(2) Diagrams with one tadpole\\
The diagrams in Figs.\ref{third-cluster}(e)-\ref{third-cluster}(l) have 
one tadpole.
If there exists a tadpole, one has to perform the integral over the tadpole 
coordinate first.
Then, the remaining gaussian integration will be of the form of 
$6 \times 6$ matrix.
Consider the diagram of Fig.\ref{third-cluster}(e).
Applying the Feynman rules, the second-order correction to 
$\Omega_0$ is given by
\begin{eqnarray}
\Omega_2^{\rm Fig.\ref{third-cluster}(e)} &=&
\sum_{s,t, u\ge 1; v \ge 0} (\pm z)^{s+t+u+v} \int_0^\beta d \beta_1 
\int_0^{\beta_1} d \beta_2 \int \left ( \prod_{i=1}^4 d{\rm\bf r}_i \right )
{\cal V}^H ({\rm\bf r}_1,{\rm\bf r}_2)
{\cal V}^H ({\rm\bf r}_3,{\rm\bf r}_4)
\nonumber \\
&& \times 
G_{s \beta}({\rm\bf r}_2,{\rm\bf r}_2)
G_{v\beta+\beta_{12}}({\rm\bf r}_1,{\rm\bf r}_3)
G_{u \beta}({\rm\bf r}_3,{\rm\bf r}_4)
G_{t\beta-\beta_{12}}({\rm\bf r}_4,{\rm\bf r}_1).
\label{omega2e}
\end{eqnarray}
Using the representation of Eq. (\ref{greensbeta}), the above reads as
\begin{eqnarray}
\Omega_2^{\rm Fig.\ref{third-cluster}(e)} &=&
\sum_{s,t, u\ge 1; v \ge 0} (\pm ze^{-x})^{s+t+u+v} \int_0^\beta d \beta_1 
\int_0^{\beta_1} d \beta_2 \int \left ( \prod_{i=1}^4 d{\rm\bf r}_i \right )
a_s a_t a_u a_v \left ( {\omega_c \over \pi} \right )^4 
\nonumber \\
&& 
\times \pi^2 | \alpha |^2 \delta({\rm\bf r}_{12}) \delta({\rm\bf r}_{34})
\exp\left \{
-2 b_s r_2^2 
- {\omega_c \over 2} a_v \left ( r_{13}^2 +
2 i \epsilon \hat {\rm\bf k} \cdot ({\rm\bf r}_1 \times {\rm\bf r}_3 )
\right ) 
- b_v ( r_1^2 + r_3^2 ) \right .
\nonumber \\ 
&&
~~~~~~~~
- {\omega_c \over 2} a_u \left ( r_{34}^2 +
2 i \epsilon \hat {\rm\bf k} \cdot ({\rm\bf r}_3 \times {\rm\bf r}_4 )
\right ) 
-b_u ( r_3^2 + r_4^2 )
\nonumber \\
&&~~~~~~~~
\left .
- {\omega_c \over 2} a_t \left ( r_{41}^2 +
2 i \epsilon \hat {\rm\bf k} \cdot ({\rm\bf r}_4 \times {\rm\bf r}_1 )
\right ) - b_t ( r_4^2 + r_1^2 ) 
\right \}.
\label{omega2ea}
\end{eqnarray}
It is easy to show that 
this leads to $\Omega_2^{\rm Fig.\ref{third-cluster}(m)}$.
In order to investigate the non-triviality of this diagram 
in connection with the Moyal phase factors,
we introduce the constant vertex of $\alpha \epsilon \omega_c /2 $ 
in (\ref{calVH}).
After integration over the tadpole coordinate ${\rm\bf r}_2$, 
we perform the gaussian integration over ${\rm\bf r}_1, {\rm\bf r}_3$ and 
${\rm\bf r}_4$.
One obtains $\pi^3/\sqrt{\det M_6}$ where the matrix $M_6$ is given by
\begin{equation}
M_6 = \left (
\begin{array}{cccccc}
c_{11} & c_{13}  & c_{14}  & 0       & d_{13}  & d_{14} \\
c_{13} & c_{33}  & c_{34}  & -d_{13} & 0       & d_{34} \\
c_{14} & c_{34}  & c_{44}  & -d_{14} & -d_{34} & 0      \\
0      & -d_{13} & -d_{14} & c_{11}  & c_{13}  & c_{14} \\
d_{13} & 0       & -d_{34} & c_{13}  & c_{33}  & c_{34} \\
d_{14} & d_{34}  & 0       & c_{14}  & c_{34}  & c_{44} 
\end{array}
\right )
\label{M6}
\end{equation}
where
\begin{eqnarray}
c_{11} &=& { \omega_c \over 2} ( a_v + a_t ) + b_v + b_t ,
\nonumber \\
c_{33} &=& { \omega_c \over 2} ( a_u + a_v ) + b_u + b_v ,
\nonumber \\
c_{44} &=& { \omega_c \over 2} ( a_t + a_u ) + b_t + b_u ,
\nonumber \\
c_{13} &=& -{ \omega_c \over 2} a_v ,
\nonumber \\
c_{14} &=& -{ \omega_c \over 2} a_t ,
\nonumber \\
c_{34} &=& -{ \omega_c \over 2} a_u ,
\nonumber \\
d_{13} &=& i \epsilon { \omega_c \over 2} a_v = -i \epsilon c_{13},
\nonumber \\
d_{14} &=& - i \epsilon { \omega_c \over 2} a_t = i \epsilon c_{14},
\nonumber \\
d_{34} &=&  i \epsilon { \omega_c \over 2} a_u = -i \epsilon c_{34}.
\nonumber 
\end{eqnarray}
Again, one finds that the determinant of $M_6$ becomes a perfect aquare:
\begin{eqnarray}
\det M_6 &=& \left\{ 
\det \left  (
\begin{array}{ccc}
c_{11} & c_{13} & c_{14} \\
c_{13} & c_{33} & c_{34} \\
c_{14} & c_{34} & c_{44} 
\end{array}
\right )
-c_{11} d_{34}^2 -c_{33} d_{14}^2 -c_{44} d_{13}^2 
\right .
\nonumber \\
&&~~~~~~
+2 c_{13} d_{34} d_{14} -2 c_{14} d_{34} d_{13} 
+ 2 c_{34} d_{14} d_{13} 
\Bigg \}^2.
\label{detM1}
\end{eqnarray}
Finally $\sqrt{\det M_6}$ takes the form
\begin{eqnarray}
\sqrt{\det M_6} &=& c_{11}c_{33}c_{44} 
- c_{11} \left ( c_{34}^2 + d_{34}^2 \right ) 
- c_{33} \left ( c_{14}^2 + d_{14}^2 \right ) 
- c_{44} \left ( c_{13}^2 + d_{13}^2 \right )
\nonumber \\ 
&&
+ 2 c_{34} \left ( c_{13}c_{14} + d_{13} d_{14} \right ) 
+ 2 d_{34} \left ( c_{13}c_{14} - d_{13} d_{14} \right ).
\label{sqrtM}
\end{eqnarray}
Here we observe that the Moyal phase factor($d_{ij}^2$) contribute to 
$\Omega_2$ as exactly opposed to the free-particle nature($c_{ij}^2$).
The last term is considered as a compositive term.

(3) Diagrams without tadpole\\
Diagrams in Figs.\ref{third-cluster}(a)-(d) and 
Figs.\ref{second-cluster}(a)-(d) belong to this category.
In these cases one finds that the integrand can be cast into the form as 
\begin{equation}
\exp \left \{
-\sum_{i,j=1}^4 c_{ij} {\rm\bf r}_i \cdot {\rm\bf r}_j
-\sum_{i,j=1}^4 d_{ij} \hat{\rm\bf k} \cdot 
   ({\rm\bf r}_i \times {\rm\bf r}_j )
\right \},
\label{inregrand1}
\end{equation}
where $c_{ij}$ involve the free-particle nature, 
whereas $d_{ij}$ contain the Moyal phase factors.
Performing the gaussian integration over 
${\rm\bf r}_1,{\rm\bf r}_2,{\rm\bf r}_3,{\rm\bf r}_4$, one obtains 
$\pi^4/\sqrt{\det M_8}$ with the $8\times 8$ matrix
\begin{equation}
M_8 = \left (
\begin{array}{cc}
c_{ij} & d_{ij} \\
-d_{ij} & c_{ij}
\end{array}
\right ).
\label{M8}
\end{equation}
Its determinant becomes a perfect square of the form $\det M = (\det P + Q)^2$
with $\det P = \det (c_{ij})$ and $ Q = Q(c_{ij}, d_{ij})$\cite{Kim93PRD4839}.
After some manipulation, $\det M$ can be lead to the similar form as in
(\ref{sqrtM}). 

\section{Discussions}
\label{discussion}
In this paper, we study the effect of Moyal phase factors on the 
thermodynamic potential using the anyonic model in the presence of a 
magnetic field.
In this case, we use the coordinate space green's function including 
the Moyal phase factor without manipulating the vertices.
It turns out that the Moyal phase factors contribute to the 
thermodynamic potential $\Omega$ as opposed to the 
free-particle nature.
Moyal phase factors are encoded in the antisymmetric 
submatrix($d_{ij}$), whereas the free-particle properties are encoded 
in the symmetric submatrix($c_{ij}, i \ne j$).
The diagonal elements of $c_{ij}$ denote the regularization scheme.

In connection with string theory, we compare our model with 
Bigatti and Susskind's case\cite{Big9908056}.
They introduced a dipole with two opposite charges and harmonic 
interaction(${k \over 2} r_{12}^2$) between them in the 
presence of the strong magnetic field\cite{She9901080}.
They also neglected the kinetic terms and introduce the interaction 
potential $V({\rm\bf r}_1) = \lambda \delta({\rm\bf r}_1)$ by hand 
to extract the Moyal phase factor.
In the quantum level, they derived the Moyal bracket phase
$e^{i p \wedge q}$ as vertex correction in the momentum space.
Here we use the $N$ particles with the same charge $q = -e$ and 
a harmonic regulating potential(${1 \over 2} \sum_{i=1}^N k r_i^2$).
Further we use the hermitian point vertex
(${\cal V}^H = {\pi \over m} | \alpha | \delta ( {\rm\bf r}_{12})$) to study 
the higher order corrections.
Then the Moyal phase factors are included in the green's function
and thus we don't need to correct the vertex.
Although our model seems not to connect with the string theory, 
this shows clearly the noncommutative effect on the 
thermodynamic potential.

\section*{Acknowledgement}
This work was supported by the Brain Korea 21
Program, Ministry of Education, Project No. D-0025.

\end{document}